\documentclass[sigconf]{acmart}
\usepackage{booktabs} %
\setcopyright{acmcopyright}
\usepackage{algorithm}
\usepackage{enumitem}
\usepackage{multirow}
\usepackage{tabularx}
\usepackage{subfigure}
\usepackage{fontenc}
\usepackage{float}
\usepackage[noend]{algpseudocode}
\newcommand{\name}{Inf-VAE}

\usepackage{amsmath,amsfonts,bm}

\def\eqref#1{equation~\ref{#1}}

\def\1{\bm{1}}

\def\rv{{\textnormal{v}}}

\def\rva{{\mathbf{a}}}
\def\rvb{{\mathbf{b}}}

\def\rvh{{\mathbf{h}}}

\def\rvp{{\mathbf{p}}}

\def\rvv{{\mathbf{v}}}

\def\rvz{{\mathbf{z}}}

\def\mA{{\bm{A}}}

\def\mD{{\bm{D}}}

\def\mI{{\bm{I}}}

\def\mL{{\bm{L}}}

\def\mP{{\bm{P}}}

\def\mV{{\bm{V}}}
\def\mW{{\bm{W}}}
\def\mX{{\bm{X}}}

\def\mZ{{\bm{Z}}}

\DeclareMathAlphabet{\mathsfit}{\encodingdefault}{\sfdefault}{m}{sl}
\SetMathAlphabet{\mathsfit}{bold}{\encodingdefault}{\sfdefault}{bx}{n}

\def\gC{{\mathcal{C}}}

\def\gE{{\mathcal{E}}}

\def\gG{{\mathcal{G}}}

\def\gL{{\mathcal{L}}}

\def\gN{{\mathcal{N}}}

\def\gU{{\mathcal{U}}}
\def\gV{{\mathcal{V}}}

\def\sD{{\mathbb{D}}}

\def\sR{{\mathbb{R}}}

\def\sT{{\mathbb{T}}}

\newcommand{\E}{\mathbb{E}}

\newcommand{\sigmoid}{\sigma}

\newcommand{\KL}{D_{\mathrm{KL}}}

\copyrightyear{2020}
\acmYear{2020}
\setcopyright{acmcopyright}
\acmConference[WSDM '20]{The Thirteenth ACM International Conference on Web Search and Data Mining}{February 3--7, 2020}{Houston, TX, USA}
\acmBooktitle{The Thirteenth ACM International Conference on Web Search and Data Mining (WSDM '20), February 3--7, 2020, Houston, TX, USA}
\acmPrice{15.00}
\acmDOI{10.1145/3336191.3371811}
\acmISBN{978-1-4503-6822-3/20/02}

\newcommand\independent{\protect\mathpalette{\protect\independenT}{\perp}}
\def\independenT#1#2{\mathrel{\rlap{$#1#2$}\mkern2mu{#1#2}}}

\begin{document}
\fancyhead{}
\title{Inf-VAE: A Variational Autoencoder Framework to Integrate Homophily and Influence in Diffusion Prediction}

\stepcounter{footnote}

\author{\vspace{-10pt}Aravind Sankar, Xinyang Zhang, Adit Krishnan, Jiawei Han}
\affiliation{
  \institution{University of Illinois at Urbana-Champaign, IL, USA}
  \{asankar3, xz43, aditk2, hanj\}@illinois.edu
}

\renewcommand{\authors}{Aravind Sankar, Xinyang Zhang, Adit Krishnan, Jiawei Han}

\renewcommand{\algorithmicrequire}{\textbf{Input:}}
\renewcommand{\algorithmicensure}{\textbf{Output:}}
\renewcommand{\citep}{\cite}

\makeatletter
\newcommand{\algmargin}{\the\ALG@thistlm}
\makeatother
\begin{abstract}
Recent years have witnessed
tremendous interest in understanding and predicting information spread on 
social media platforms such as Twitter, Facebook, etc.
Existing diffusion prediction methods primarily exploit the sequential order of influenced users by projecting diffusion cascades onto their local social neighborhoods.
However, this fails to capture global social structures that do not explicitly manifest in any of the cascades, resulting in poor performance for inactive users with limited historical activities.

In this paper, we present a novel variational autoencoder framework (\name) to jointly embed
\textit{homophily} and \textit{influence} through proximity-preserving \emph{social} and position-encoded \emph{temporal} latent variables. To model social homophily, Inf-VAE utilizes powerful graph neural network architectures to learn social variables that selectively exploit the social connections of users.
Given a sequence of seed user activations, Inf-VAE uses a novel expressive \textit{co-attentive fusion network} that jointly attends over their social and temporal variables to predict the set of all influenced users. 
Our experimental results on multiple real-world social network datasets, including Digg, Weibo, and Stack-Exchanges demonstrate significant gains (22\% MAP$@10$) for~\name~over state-of-the-art diffusion prediction models; we achieve massive gains for users with sparse activities, and users who lack direct social neighbors in seed sets.

\end{abstract}


\maketitle

\section{Introduction}
\label{sec:intro}
In social media, information disseminates or \emph{diffuses} to a large number of users through posting or re-sharing behavior, resulting in a \textit{cascade} of user activations, \textit{e.g.}, a user voting a news story on Digg (a social news sharing website) triggers a series of votes from multiple users, who may be his friends or other users interested in the same story.
Given a set of \textit{activated} seed users, diffusion models aim to predict the set of all influenced users.
Diffusion modeling has widespread social media applications, including viral marketing~\cite{viral}, recommendations~\cite{recom, longtail}, and popularity prediction~\cite{popularity}.

The diffusion prediction problem has received significant attention in the research community.
Unlike pre-defined propagation hypotheses~\cite{ic}, recent methods learn data-driven diffusion models from collections of user activation sequences (\textit{diffusion cascades}).
Existing diffusion models broadly fall into two categories.

\textit{Probabilistic generative cascade} models use hand-crafted features including roles~\cite{rain}, communities~\cite{cic-icdm}, topics~\cite{topic-ic}, and structural patterns~\cite{structinf}. Such methods
rely on feature engineering that requires manual effort and extensive domain knowledge, and are limited by the modeling capacity of carefully chosen probability distributions.

\textit{Representation learning} methods avoid feature extraction by learning user embeddings characterizing their influencing ability and conformity~\cite{wsdm16,inf2vec}.
State-of-the-art methods project cascades onto local social neighborhoods to generate Directed Acyclic Graphs (DAGs), and propose extensions of Recurrent Neural Networks (RNNs). %
In particular, DAG-structured LSTMs~\cite{topolstm} explicitly operate on the induced DAG, while attention-based RNNs~\cite{cyanrnn,deepdiffuse,cikm18_attention} implicitly consider cross-dependence for diffusion prediction.

Prior works only consider the sequence or projected social structure (induced DAG) of previously influenced users while ignoring \textit{social structures that do not manifest in cascades}. As a result, they only capture the temporal correlation of diffusion behaviors among users, which is also known as \textit{temporal influence} or \textit{contagion}~\cite{shalizi}.
Consider a Twitter user with interests in politics, who is likely to follow famous political leaders and join interest groups that induce transitive connections to other users; however, these connections may not appear in cascades unless she re-tweets or posts content. 
\textit{Social homophily}~\cite{homophily} suggests that ties are more likely between users with shared traits or interests, which can induce correlated diffusion behaviors without direct causal influence.
Since a vast majority of social media users seldom post content and thus rarely appear in cascades, it is critical to exploit their social neighborhood structures to characterize social homophily accurately.

However, homophilous diffusion and contagion can result in significantly different dynamics, \textit{e.g.}, contagions are self-reinforcing and viral while homophily hinges on users' preferences or traits.
Real-world cascades are often a complex combination of both aspects with user-specific variations.
Indeed, it is well known that social homophily and temporal influence are fundamentally confounded in observational studies~\cite{shalizi}.
Thus, we propose a data-driven framework to contextually model their joint effect when predicting user-level diffusion behaviors.
Therefore, our key objective is to \textit{develop a principled neural framework to unify social homophily and temporal influence in diffusion prediction}.

Our architecture~\name~jointly models \emph{homophily} through \emph{social} embeddings preserving social network proximity and \emph{influence} through \emph{temporal} embeddings encoding the relative sequential order of user activations.
Motivated by the recent successes of variational autoencoders (VAEs)~\cite{vae} in characterizing sparse users via
Gaussian priors~\cite{vaecf}, and the expressive power of graph neural networks~\cite{gcn, graph_enc_dec}, we adopt VAEs to model social homophily.
We learn structure-preserving social embeddings through a VAE framework that supports a wide range of graph neural network architectures as encoders and decoders.
Given an initial set of seed user activations,~\name~utilizes an expressive \textit{co-attentive fusion network} that captures complex non-linear correlations between social and temporal embeddings, to model their joint effect on predicting the set of all influenced users. We make the following contributions:

\begin{itemize}[leftmargin=*]

\item \textbf{Generalizable Variational Autoencoder Framework}:
Unlike existing diffusion prediction methods that only consider local induced propagation structures,~\name~is a generalizable VAE framework that models social homophily through graph neural network architectures of arbitrary complexity, to selectively exploit the rich global network of social connections.

\item \textbf{Efficient Homophily and Influence Integration:}
To the best of our knowledge,  ours is the first work to comprehensively exploit social homophily and temporal influence in diffusion prediction. Given a sequence of seed user activations,~\name~employs an expressive \textit{co-attentive fusion network} to jointly attend over their social and temporal embeddings to predict the set of all influenced users.
~\name~with co-attentions is faster than state-of-the-art recurrent methods by an order of magnitude.

\item \textbf{Robust Experimental Results:}
Our experiments on multiple real-world social networks, including Digg, Weibo, and Stack-Exchanges, demonstrate significant gains for~\name~over state-of-the-art models.
Modeling social homophily through VAEs enables massive gains for users with \textit{sparse activities}, and users who \textit{lack direct social neighbors in seed sets}. An ablation analysis of
various modeling choices further highlights the synergistic effects of jointly modeling homophily and temporal influence.

\end{itemize}

\section{Related Work}
\label{sec:related}
We discuss existing work on diffusion modeling followed by
related work on network representation learning and co-attentions.

\textbf{Information diffusion overview.}
Historically, information diffusion has been studied through two seminal models: Independent Cascade (IC)~\citep{ic} and Linear Threshold (LT)~\citep{lt}.
Three distinct applications emerged, namely: \textit{network inference}~\cite{netinf}, which infers the underlying social network that best explains the observed cascades; \textit{cascade prediction}~\cite{deepcas}, which predicts macroscopic properties of cascades, including size, growth, and shape; and \textit{diffusion prediction}~\cite{topolstm}, which learns a model from social links and cascade sequences, to predict the set of influenced users given a seed set of activated users.
In this paper, we focus on diffusion prediction.

\textbf{Diffusion prediction.} 
The earliest data-driven methods propose several extensions of IC and LT incorporating topics~\cite{topic-ic}, continuous timestamps~\cite{ctic}, user profiles~\cite{node_attribute}, and community structure~\cite{cic-icdm}. A few techniques explore probabilistic generative models via latent topics and communities~\citep{cold,hcid}. 
Most recent studies focus on learning representations to overcome feature engineering or pre-defined hypotheses in diffusion modeling~\cite{cdk,aaai15,wsdm16,topolstm,cyanrnn,deepinf,inf2vec,cikm18_attention}.
Emb-IC~\cite{wsdm16}, Inf2vec~\cite{inf2vec} embed user influencing capability and susceptibility in diffusion.
Topo-LSTM~\cite{topolstm}, CYAN-RNN~\cite{cyanrnn}, SNIDSA~\cite{cikm18_attention}, and DeepDiffuse~\cite{deepdiffuse} project the diffusion cascades on local social neighborhoods and model the resulting DAG propagation structures with RNNs.
These techniques outperform classical approaches by significant margins in diffusion prediction. 
Our key observation is that these projected DAGs could ignore social structures that do not appear in any observed cascade. In contrast, our model~\name~can account for unobserved social connections in the user activation process by modeling social homophily through VAEs.

A related problem is social influence prediction, which aims to classify social media users based on the activation status of their ego-network~\cite{locality, deepinf}.
Direct extensions to predict the set of all influenced users (diffusion prediction) entails reapplying their models on each candidate inactive user in the social network, resulting in prohibitive inference costs, hence preventing a comparison. 

\textbf{Network representation learning:}
This line of work captures varied notions of structural node proximity~\cite{node2vec, rase} in networks via low-dimensional vectors.
Notably, graph neural networks have achieved great success
in node classification and link prediction~\cite{gcn,graphsage,gat, motifcnn, metagnn, dysat_arxiv, Narang2019induced}.
Graph Autoencoders~\cite{sdne,vgae} employ various encoding and decoding architectures to embed network structure and learn unsupervised node embeddings.
~\citet{graph_enc_dec} unify a large family of network embedding methods
in an autoencoder framework.
However, general-purpose embeddings modeling structural proximity are not directly suited to diffusion modeling.

\textbf{Co-attentional models:}
Our work also leverages recent advances in neural attention mechanisms, especially in Natural Language Processing~\cite{attention}. 
Specifically, co-attention has achieved great success in modeling relationships between pairs of sequences, \textit{e.g.}, question-answer~\cite{dynamic_coattention}, etc.
Co-attentional methods compute interaction weights between data modalities, learning fine-grained non-linear correlations. In our work, we develop a co-attentive fusion network to capture the contextual interplay of users' social and temporal representations for diffusion prediction.

\section{Problem Definition}
\label{sec:defn}
We study diffusion prediction where the goal is to predict the set of all influenced users, given temporally ordered seed user activations.
\begin{definition}\textbf{Social Network:} The social network is represented as a graph $\gG = (\gV,\gE)$ where $\gV = \{ v_i\}_{i=1}^N$ is the set of $N$ users and $\gE = \{ e_{ij} \}_{i,j=1}^N$ is the set of links. We denote the adjacency matrix of $\gG$ by $A \in \sR^{N \times N}$ where $A_{i,j} = 1$ if $e_{i,j} \in \gE$ otherwise 0.
\end{definition}
\vspace{-8pt}
\begin{definition}\textbf{Diffusion cascade:} A diffusion cascade $D_i$ is an ordered sequence of user activations in ascending order of time denoted by: $D_i = \{ (v_{i_k}, t_k) \mid v_{i_k} \in \gV, t_k \in [0, \infty), \; k = 1\dots K\}$, each $v_{i_k}$ is a distinct user in $\gV$ (no repeats) 
and $t_k$ is non-decreasing, \textit{i.e.}, $t_k \leq t_{k+1}$. The $k^{th}$ user activation is recorded as tuple $(v_{i_k}, t_k)$, referring the activated user and activation time. 
\end{definition}
\vspace{-3pt}
We represent cascades by delay-agnostic relative activation orders similar to~\cite{inf2vec,topolstm,wsdm16}, \textit{i.e.}, a cascade is equivalently written as $D = \{(v_{i_k} , k) \mid v_{i_k} \in \gV\}_{k=1}^K$. We do not assume the availability of explicit re-share links between users in cascades; this corresponds to the simplest yet most general setting of diffusion~\cite{topolstm,inf2vec}.
Though timestamps may be easily used as input features, we leave generation of continuous timestamps as future work.
\vspace{-5pt}
\begin{definition} \textbf{Diffusion prediction:}
Given a social network $G$ and a collection of cascade sequences $\sD = \{D_i, 1 \leq i \leq |\sD|\}$, learn diffusion model $M$ to predict the future set of influenced users in a cascade with the seed activation sequence $I = \{(v_{i_1}, 1), \dots, (v_{i_k},k) \}$ of $k$ seed users. %
Diffusion prediction estimates the probability of influencing each inactive user: $P_\Theta(v \mid I) \; \forall v \in \gV -  I$, inducing a ranking of activation likelihoods over the set of inactive users. \vspace{-3pt}
\end{definition}
\vspace{-3pt}
We create a training set $\sT$ of diffusion \emph{episodes} containing (seed activations, activated users) tuples from the cascade collection $\sD$, by randomly splitting each cascade $D \in \sD$ of length $K$ at each time step $ 2 \leq k \leq K-1$.
Specifically, a split at time step $k \geq 2$, creates a training episode $(I_{k}, C_{k})$ where $I_k = \{ (v_{i_j}, j);  1 \leq j \leq k \}$ is the seed set consisting of the cascade sliced at $k$ and $C_{k} = \{ v_{i_{k+1}}, \dots, v_{i_K} \}$ is the set of influenced users after time step $k$. Thus, we denote the training set by $\sT = \{ (I_i, C_i ) \; 1 \leq i \leq |\sT| \}$.

\section{Influence Variational Autoencoder}
\label{sec:model}
In this section, we describe our proposed Influence Variational  Autoencoder (\name) for predicting information diffusion.

\subsection{Model Description}
We describe the latent variables modeling social homophily and temporal influence, followed by our generative network~\name.
\subsubsection{\textbf{Social Homophily.}}
\label{sec:structural}
Our objective is to define latent \textit{social variables} for users that capture social homophily.
The homophily principle stipulates that users with similar interests are more likely to be connected. In the absence of explicit user attributes, we posit that 
highly interconnected users in social communities share homophilous relationships.
We model social homophily through latent \textit{social} variables designed to encourage users with shared social neighborhoods to have similar latent representations.

Specifically, we assign a latent \emph{social} variable $\rvz_i$ for user $v_i$, where the prior for $\rvz_i$ is chosen to be a unit normal distribution, in line with standard assumptions in VAEs.
Normal distributions are chosen in VAE frameworks due to their flexibility to support arbitrary functional parameterizations by isolating sampling stochasticity to facilitate back-propagation~\cite{vae}.
We assume the latent social variables $\mZ$ to collectively generate the social network $\gG$, through a graph generation neural network $f_{\textsc{dec}} (\mZ)$ parameterized by $\theta$. The corresponding generative process is given by:
\begin{equation}
\rvz_i \sim \gN (0, I_D) \hspace{10pt} \gG \sim p_{\theta}(\gG \mid \mZ) =  p_{\theta}(\gG \mid  f_{\textsc{dec}} (\mZ))
\label{eqn:struct_gen}
\end{equation}
where $I_D \in \sR^{D \times D}$ is an identity matrix of $D$ dimensions. Here, the graph generation neural network $f_{\textsc{dec}} (\mZ)$ can be instantiated to preserve an arbitrary notion of structural proximity in the social network $\gG$ (Sec~\ref{sec:neural_vgae}).
In the above equation, we abuse the notation of $\gG$ to denote an appropriate representational form of the social network structure, which can take multiple forms, including the adjacency matrix, random walks sampled from $\gG$, etc.

While homophily characterizes peer-to-peer interest similarity, its impact on user behaviors tends to asymmetric since users who share interests may drastically differ in their posting rates, \textit{e.g.}, certain users are naturally predisposed to be socially active and hence more \textit{influential} in comparison to others.
Thus, it is necessary to differentiate user \textit{roles} when modeling the effect of social homophily on diffusion behaviors.
Similar concepts have been examined in social influence literature to characterize users by their influencing capability and conformity~\cite{conformity,aaai15,wsdm16,topolstm,inf2vec}.

We associate each user $v_i \in \gV$ with a \textit{sender} $\rvv^{s}_i \in \sR^{D}$ and \textit{receiver} $\rvv^{r}_i \in \sR^{D}$ latent variable.
Our key innovation lies in conditioning the information sending and receiving capabilities of users on their homophilous traits.
We use normal distributions centered at $\rvz_i$ 
to define the \emph{sender} and \emph{receiver} variables for user $v_i$ as: 
\begin{align}
\rvv^{s}_i &\sim \gN (\mathbf{z}_i, \lambda_s^{-1} I_D) \hspace{10pt} \rvv^{r}_i \sim \gN (\mathbf{z}_i , \lambda_r^{-1} I_D) 
\end{align}
where $\lambda_s,\lambda_r$ are hyper-parameters controlling the degree of variation or uncertainty for $\rvv^{s}_i$ and $\rvv^{r}_i$ \textit{w.r.t.} $\rvz_i$.
Let $V_S$ and $V_R $ denote the set of all sender and receiver variables respectively for all users.
\begin{figure*}[t]
    \centering
    \includegraphics[width=0.9\linewidth]{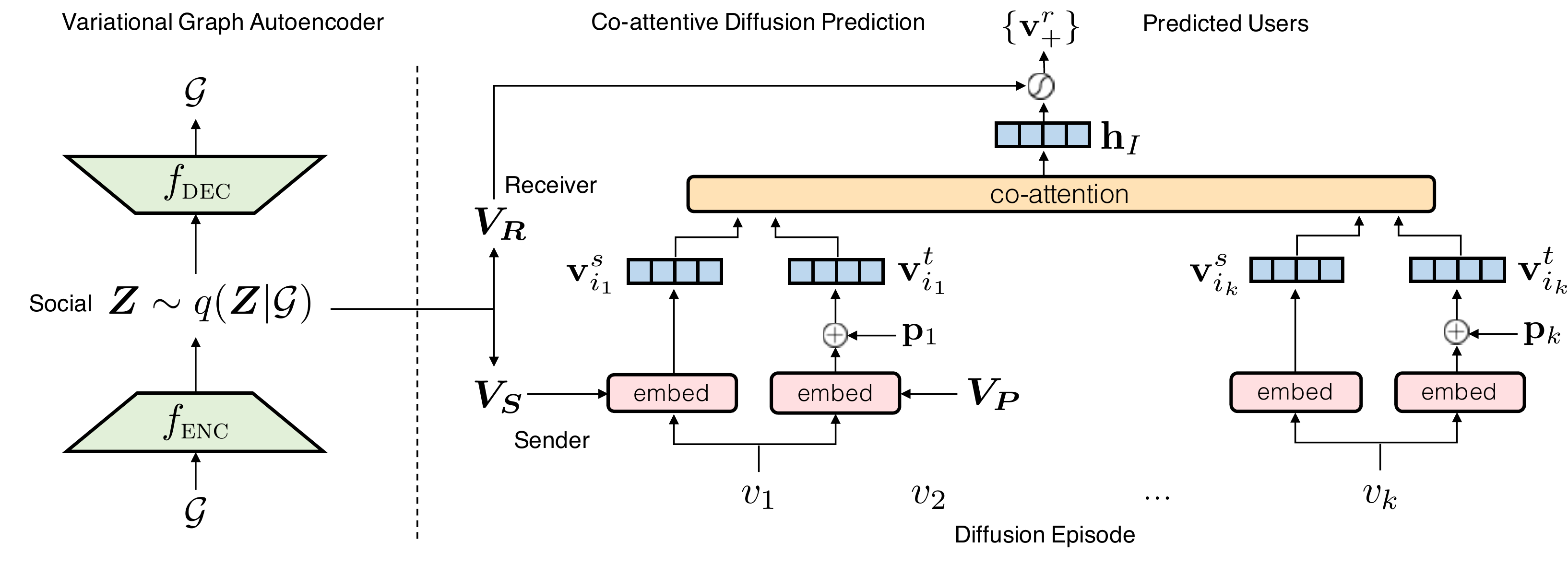}
    \vspace{-14pt}
    \caption{Neural Architecture of~\name~depicting latent variable interactions. The left side indicates the VAE framework to model social homophily; right side denotes the co-attentive fusion network to integrate the social and temporal variables.}
    \label{fig:arch}
    \vspace{-10pt}
\end{figure*}

\begin{table}[t]
\small
    \centering
    \begin{tabular}{@{}c|l@{}}
    \toprule
        Symbol &  Description \\
        \midrule
         $\mZ$ &  Social variables modeling network proximity, for all users $\gV$ \\
         $\mV_S$ &  Sender variables for all users $\gV$ \\
         $\mV_R$ &  Receiver variables for all users $\gV$ \\
         $\mV_{T}$ & Temporal influence variables for all users $\gV$ \\
         $\mV_{P}$ & User-specific popularity variables for all users $\gV$ \\
         $\mP_K$ & Position-encoded temporal embeddings for all time steps $K$ \\
         \bottomrule
    \end{tabular}
    \caption{Notations}
    \label{tab:notations}
    \vspace{-20pt}
\end{table}
\subsubsection{\textbf{Temporal Influence.}} Now, we define latent \textit{temporal influence} variables to describe the varying influence effects of seed users depending on the relative sequential order of activations.
There are two interesting factors at play: activation orders and popularity effects.
A majority of social media users adopt more recent information while often ignoring old and obsolete content~\cite{implicit_diffusion}.
On the other hand, social status impacts the influencing power of seed users independent of their activation order and social neighbors,
\textit{e.g.}, famous media figures naturally exert significant influence.
Thus, we consider both the relative sequential order of user activations and popularity effects of seed users to model temporal influence.

To quantify the temporal influence exerted by a seed user activation $(v_{i_k}, k)$ of user $v_{i_k}$ at time step $k$ ($ 1\leq k \leq K$), we first encode the relative position $k$ through positional-encodings~\cite{posn-enc} to obtain temporal embeddings $\rvp_k$.
Since we expect the variation in popularity effects to be quite small, we draw
user-specific popularity variables from a zero-mean normal distribution to serve as offsets to the temporal embeddings.
Specifically, the \emph{temporal influence} variable for activation $(v_{i_k}, k)$ denoted by $\rvv^{t}_{i_k}$, is given by:
\vspace{-3pt}
\begin{align}
    \rvv^{p}_{i_k} \sim \gN (0, \lambda_{p}^{-1} I_D) \hspace{15pt}  \rvp_k = PE(k) \hspace{15pt} \rvv^{t}_{i_k} =  \rvv^{p}_{i_k} + \rvp_k \\
    PE(k)_{2d} = sin(k/10000^{2d/D}) \hspace{5pt} PE(k)_{2d+1} = cos(k/10000^{2d/D}) \nonumber
\end{align}
where $\lambda_{p}$ is a hyper-parameter to control the popularity effects, and $1 \leq d \leq D/2$ denotes the dimension in the temporal embedding $\rvp_k$.
Note that the popularity variable $\rvv^{p}_{i_k}$ is user-specific, while temporal embedding $\rvp_k$ only depends on the activation step $k$.
The set of all latent user popularity variables are denoted by $\mV_{P}$, while $\mP_K$ represents the set of position-encoded \emph{temporal} embeddings.
\subsubsection{\textbf{Co-attentive Diffusion Episode Generation.}}
Let us consider a single diffusion episode $(I, C) \in \sT$, with seed activations $I = \{ (v_{i_1}, 1), \dots, (v_{i_k}, k)\}$ and influenced users $C = \{ v_{i_{k+1}}, \dots, v_{i_K}\}$.
A diffusion model aims to predict the set of influenced users $C$ given seed activations $I$.
Since diffusion is always conditioned on $I$, we propose a conditional generative process to sample $C$ given $I$. %

Let us denote the set of seed users by $I_U = \{ v_{i_1}, \dots, v_{i_k} \}$.
Our objective is to jointly model the effects of social homophily and temporal influence exerted by seed users $I_U$, which can be summarized by:
\textit{sender sequence} $(\rv^s_{i_1}, \rv^s_{i_2}, \dots, \rv^s_{i_K})$; and \textit{temporal influence sequence} $(\rv^{t}_{i_1}, \rv^{t}_{i_2}, \dots, \rv^{t}_{i_K})$.
To model complex correlations between the sender and temporal influence sequences, 
we propose an expressive \textit{co-attentive} fusion strategy to learn attention scores for each seed user by modeling interactions between the two sequences.
We describe the conditional generative process in two steps:

\vspace{-2pt}
\begin{itemize}[leftmargin=*]
\item The social homophily and temporal influence aspects of seed users, are integrated into an aggregate seed set representation $\rvh_I$.
The co-attentive fusion network $G_{\textsc{diff}} (\cdot)$ performs homophily-guided temporal attention, \textit{i.e.}, attends over the temporal influence variables by computing co-attentional weights that jointly depend on both homophily and temporal influence characteristics.
As illustrated in Figure~\ref{fig:arch}, the sender and temporal influence variables of seed users feed into a fusion network $G_{\textsc{diff}}(\rvv^s_{i_k}, \rvv^{t}_{i_k})$.
The aggregate seed representation $\rvh_I$ is computed as:
\begin{align}
\alpha_{k} = \frac{ \exp( G_{\textsc{diff}}(\rvv^s_{i_k}, \rvv^{t}_{i_k} (k))) }{\sum\limits_{j=1}^K \exp(G_{\textsc{diff}}(\rv^s_{i_j}, \rv^{t}_{i_j} (j))) }  \hspace{10pt} \rvh_I  = \sum\limits_{j=1}^K \alpha_j \rvv^{t}_{i_j} (j)
\end{align}
Each $\alpha_j$ is the normalized co-attentional coefficient for seed user $v_{i_k}$ denoting its  contribution in computing the aggregate representation $\rvh_I$.
To model the co-dependence between $\rvv^s_i$, $\rvv^{t}_i$, we define the co-attentive function $G_{\textsc{diff}}(\rvv^s_{i_k}, \rvv^{t}_{i_k}) = tanh({\rvv^s_{i_k}}^T \mW \rvv^{t}_{i_k})$ as a bi-linear product parameterized by $\mW \in \sR^{D \times D}$. %

\item The probability of influencing an inactive user $v_j$ depends on the sending capacity of seed users (embedded in $\rvh_I$) and her receiving capability (encoded by \emph{receiver} variable $\rvv_j^r$).
We quantify the likelihood of influencing $v_j$ by $\rvh_{I}^T \rvv_j^r$.
For each inactive user $v_j \in \gV- I_U$, we draw a binary variable $\gC_j \in \{0,1\}$ indicating whether user $v_j$ is influenced by set users $I_U$ or not, given by:
\vspace{-3pt}
\begin{equation}
\gC_j \sim %
Ber (\sigma(\rvh_{I}^T \rvv_j^r)) \; \forall v_j \in \gV- \{ v_{i_1}, \dots, v_{i_K}\}
\label{eqn:diffusion_gen}
\end{equation}
\noindent where $\sigma(\cdot)$ is the sigmoid function and $Ber(\cdot)$ is the Bernoulli distribution.
The corresponding logistic log-likelihood of generating a single diffusion episode $(I, C)$ is given by:
\vspace{-2pt}
\begin{align}
\gL^{\textsc{diff}}_{I,C} &= \log p_{\theta} (C \mid I, \mV_S, \mV_R, \mV_{P})  \\ \nonumber \vspace{-3pt}
&=  \sum\limits_{v \in C} \eta \log (\sigma(\rvh_{I}^T \rvv_i^r)) +  \sum\limits_{v_n \in \gV - C - I_U} \log (1-  \sigma(\rvh_{I}^T \rvv_n^r ))
\end{align}
Here, $\eta$ re-weights positive examples since the actual number of influenced users is much smaller than the total number of users.
\end{itemize}

\subsection{Model Likelihood}
\label{sec:likelihood}

Due to the intractability of analytically computing the latent posterior distribution $p (\mV_S, \mV_R, \mV_{P}, \mZ | \gG, \sT)$, we use variational inference to factorize the posterior with a mean-field approximation:
\vspace{-3pt}
\begin{align}
q(\mV_S, \mV_R, \mV_P, \mZ | \gG) =   q(\mV_S ) q(\mV_R) q(\mV_P) q(\mZ | \gG)
\label{eqn:mean_field}
\end{align}

The variational distributions of variables $\mV_S, \mV_R,$ and $\mV_P$ follow normal distributions while the 
social variables $\mZ$ are conditioned on $\gG$ through a structure-encoding inference network~\cite{vae}.
Specifically, the variational distribution of $\mZ$ denoted by $q_{\phi}(\mZ | \gG)$, is a diagonal normal distribution parameterized by $f_{\textsc{enc}} (\gG)$ defined as:
\vspace{-2pt}
\begin{equation*}
f_{\textsc{enc}} (\gG) \equiv [\mu_{\phi}(\gG),  \log \sigma_{\phi}^2 (\gG)]  \hspace{5pt} q_{\phi} (\mZ | \gG) = \gN \left(\mu_{\phi}(\gG), diag( \sigma_{\phi}^2 (\gG))\right)
\end{equation*}
The inference network outputs the parameters, $\mu_{\phi}(\gG), \sigma_{\phi} (\gG)$ of the variational distribution $q_{\phi} (\mZ | \gG)$, which is designed to approximate the corresponding posterior $p(\mZ | \gG)$.
The inference network $f_{\textsc{enc}} (\gG)$
endows the model with added flexibility to incorporate arbitrary neighborhood aggregation functions such as graph convolutions~\cite{gcn}, attentions~\cite{gat}, etc.
The variational structure distribution $q_{\phi} (\mZ | \gG)$  and the structure generative model $p_{\theta} (\gG | \mZ)$ (Eqn.~\ref{eqn:struct_gen}) together constitutes a \emph{variational graph autoencoder}~\cite{vgae}.

\subsection{Neural Graph Autoencoder Details}
\label{sec:neural_vgae}
In this section, we describe functions $f_{\textsc{enc}} (\gG)$ and $f_{\textsc{dec}} (\mZ)$ which describe the graph structure inference and generative networks of~\name.
The \emph{encoder} summarizes local social
neighborhoods into latent vectors, which are subsequently transformed by the \emph{decoder} into high-dimensional structural information (\textit{e.g.}, adjacency matrix).
~\citet{graph_enc_dec} present an encoder-decoder framework to conceptually unify a large family of graph embedding methods.
Encoder architectures fall into three major categories: embedding lookups~\cite{deepwalk,node2vec}, neighborhood vector encoding~\cite{sdne}, and neighborhood aggregation~\cite{graphsage}, while decoders comprise unary and pairwise variants.
In~\name, we explore two representative choices:

\begin{itemize}[leftmargin=*]
\item \textbf{MLP + MLP}: We use a Multi-Layer Perceptron (MLP) to both encode and decode the laplacian matrix of $\gG$, given by $\mL = \mD^{-1/2} \mA \mD^{-1/2}$. %
The neighborhood vector for user $v_i$, denoted by $\rva_i$, is the $i^{th}$ row of $\mL =  [\rva_1, \dots, \rva_N ]^T$.
The encoder is an MLP network $f_{\textsc{enc}} (\rva_i)$ which encodes $\rva_i$ into $\rvz_i$, while the decoder $f_{\textsc{dec}} (\rvz_i)$ strives to reconstruct $\rva_i$ from $\rvz_i$.
We introduce a re-weighting vector $\rvb_i = \{ b_{ij} \}_{j=1}^N$ where $b_{ij} = 1$ if $L_{ij} =0$ and $b_{ij} = \beta >1$ when $L_{ij} >0 $.
$\beta$ is a confidence parameter that re-weights the positive terms ($L_{ij} > 0$) to balance the unobserved $0's$ which far outnumber the observed links in real-world networks.
The generative process to obtain $\rva_i$ from $\rvz_i$ is given by:
\vspace{-2pt}
\begin{equation*}
\rva_i \sim p_\theta (\rva_i | \rvz_i) = \gN (f_{\textsc{dec}} (\rvz_i),  diag(\rvb_i))
\end{equation*}
where $diag(\rvb_i)$ is a diagonal matrix with non-zero entries from vector $b_i$.
The corresponding Gaussian log-likelihood is given by:
\vspace{-2pt}
\begin{equation*}
\log p_{\theta} (\mA | \mZ) = \sum\limits_{i=1}^N \log p_{\theta} (\rva_i | \rvz_i)  =  \sum\limits_{i=1}^N \big\Vert \rvb_i \odot (\rva_i - f_{\textsc{dec}}(\rvz_i))\big\Vert^2 
\end{equation*}
\item \textbf{GCN + Inner Product}: We use a Graph Convolutional Network (GCN) as the encoder and an inner product decoder that maps pairs of user embeddings to a binary indicator of link existence in $\gG$. The GCN network comprises multiple stacked graph convolutional layers to extract features from higher-order structural neighborhoods.
The input to a layer is 
a user feature (or embedding) matrix $X \in \sR^{N \times F}$ and a normalized adjacency matrix $\hat{\mA}$, where each GCN layer computes the function:
\vspace{-2pt}
$$f_{\textsc{enc}} (\mA) = \sigma(\hat{\mA} \mX \mW) \hspace{10pt} \hat{\mA} = \mD^{-1/2} \mA \mD^{-1/2} + \mI_N$$

where $\mX$ is an identity matrix encoding user identities. Each entry $A_{ij}$ of adjacency matrix $\mA$ is generated according to:
\vspace{-2pt}
\begin{equation*}
A_{ij} \sim p_\theta (A_{ij} | \rvz_i, \rvz_j) = Ber(\sigma(\rvz_i^T \rvz_j))
\end{equation*}
Similar to above, we re-weight the positive entries of $\mA$ with a confidence parameter $\beta$. The logistic log-likelihood is given by:
\vspace{-2pt}
\begin{equation*}
\log p_{\theta} (\mA | \mZ) = \sum\limits_{(i,j) \in \gE} \beta \log (\sigma(\rvz_i^T \rvz_j)) + \sum\limits_{(i,j) \notin \gE} \log (1 - \sigma(\rvz_i^T \rvz_j))
\end{equation*}
As an alternative to re-weighting positive entries, negative sampling~\cite{deepwalk} can  scale this objective to large-scale networks.
\end{itemize}

\algnewcommand{\LeftComment}[1]{\hspace{\algorithmicindent} \Statex \(\triangleright\) #1}
\algnewcommand{\parState}[1]{\State%
  \parbox[t]{\dimexpr\linewidth-\algmargin}{\strut #1\strut}}

\begin{algorithm}[t]
\caption{\name~ training with block coordinate ascent.}
\begin{algorithmic}[1]
\Require Social Network ($\gG$), Training episodes ($\sT$)
\Ensure MAP estimates of $\mV_S, \mV_R, \mV_{P}$ and parameters $\theta, \phi$.
\State Initialize latent variables from a standard normal distribution.
\State \textbf{Pre-training}: Train $f_{\textsc{dec}}(\gG|\mZ)$ and $f_{\textsc{enc}}(\mZ| \gG)$ on $\gG$ using a VAE with log-likelihood:
$$L^{\textsc{VAE}} =  \E_{\scriptscriptstyle q_{\phi}(\mZ | \gG)} \log p_\theta (\gG | \mZ) - \KL{(q_{\phi} (\mZ | \gG), p(\mZ))}$$
\While{\textit{not converged}}
\LeftComment \textbf{\textit{Optimize over social network $\gG$ }}
\For{each batch of users $\gU \subseteq \gV$}
\parState {Fix $\mV_S, \mV_R, \mV_{P}, G_{\textsc{diff}} (\cdot)$ and update weights of
$f_{\textsc{enc}} (\gG)$ and $f_{\textsc{dec}} (\mZ)$ using mini-batch gradient ascent (Eqn.~\ref{eqn:objective})}
\EndFor
\LeftComment \textbf{\textit{Optimize over diffusion episodes $\sT$ }}
\For {each batch of diffusion episodes $B\subseteq \sT$}
\parState {Fix $\mZ$, $f_{\textsc{enc}} (\gG)$, $f_{\textsc{dec}}(\mZ)$ and update $\mV_S, \mV_R, \mV_{P}$, and $G_{\textsc{diff}} (.)$ using mini-batch gradient ascent. (Eqn.~\ref{eqn:objective})}
\EndFor
\EndWhile
\end{algorithmic}
\label{alg:opt}
\end{algorithm}

\subsection{Model Inference}
The overall objective maximizes a lower bound on the marginal log likelihood, also named evidence lower bound (ELBO)~\cite{variational_inf}, given by:
\vspace{-3pt}
\begin{align}
L_q &= \E_{q} [ \log p(\gG,\sT, \mV_S, \mV_R , \mV_{P}, \mZ) -   \log q(\mV_S, \mV_R, \mV_P, \mZ | \gG)] 
\label{eqn:likelihood}
\end{align}

Note that $L_q$ is a function of both generative ($\theta$) and variational ($\phi$) parameters.
However, an analytical computation of the expectation with respect to $q_\phi(\mZ |\gG)$ is intractable, while Monte Carlo sampling prevents gradient back-propagation to the neural parameters of $f_{\textsc{enc}}(\gG)$. 
With the reparametrization trick~\cite{vae}, we instead sample $\mathbf{\epsilon} \sim \gN(0, I_{N \times D})$ and form samples of $\mZ = \mu_{\phi} (\gG)  + \epsilon \odot \sigma_{\phi} (\gG)$.
This isolates the stochasticity during sampling and the gradient with respect to $\phi$ can be back-propagated through the sampled $\mZ$.

\subsubsection{\textbf{Optimization}}
Since bayesian methods to infer latent posteriors incur high computational costs, and considering our goal of making good predictions rather than explanations, we resort to MAP (Maximum A Posteriori) estimation.
Thus, we sample $\mZ$ from $q_{\phi}(\mZ|\gG)$ using point estimates for $\mV_S, \mV_R$ and $\mV_{P}$.
We maximize the joint log-likelihood with MAP estimates of latent variables $\mV_S, \mV_R, \mV_{P}$, inference and generative network parameters $\theta,\phi$, and observations $\sT$ and $\gG$, given hyper-parameters $\lambda_s, \lambda_r, \lambda_p$:
\vspace{-3pt}
\begin{align}
\gL^{ \textsc{MAP}} &= \E_{\scriptscriptstyle q_{\phi}}  [ \log p_\theta (\gG | \mZ)] - \KL{(q_{\phi}, p(\mZ))} + \sum\limits_{\scriptscriptstyle (I,C) \in \sT} \gL^{\textsc{diff}}_{I,C} \label{eqn:objective} \\ \vspace{-3pt}
&- \sum\limits_{i=1}^N  \left( \frac{\lambda_s}{2} \E_{\scriptscriptstyle q_{\phi}} \Vert \rvv^s_i - \rvz_i \Vert^2 + \frac{\lambda_r}{2} \E_{\scriptscriptstyle q_{\phi}} \Vert \rvv^r_i - \rvz_i \Vert^2 + \frac{\lambda_p}{2} \Vert \rvv^p_i \Vert^2 \right) \notag
\end{align}

where $q_\phi$ is a shorthand for $q_{\phi}(\mZ | \gG)$, and $\E_{\scriptscriptstyle q_{\phi}(\mZ | \gG)} [\mZ]$ is equal to $\mu_\phi(\gG)$ output by the inference network.
To optimize this objective, we employ block coordinate ascent with two sets of variables, $\{ f_{\textsc{enc}}(G), f_{\textsc{dec}} (\mZ) \}$ and $\{ \mV_S, \mV_R, \mV_{P}, G_{\textsc{diff}}\}$. 
As illustrated in Alg~\ref{alg:opt}, each iteration of the algorithm proceeds in two steps, by alternating optimization over the social network and diffusion cascades.

\subsubsection{\textbf{Diffusion Prediction}}
After learning the (locally) optimal model parameters and MAP estimates of latent variables, the likelihood of influencing user $v_j$ given seed activations $I$ is given by:
\vspace{-3pt}
\begin{equation}
p(v_j | I) =  \sigmoid( h_I^T \rvv^r_j)
\label{eqn:user_prediction}
\end{equation}

\subsubsection{\textbf{Complexity}}
The cost per iteration comprises two parts:
(a) optimizing over social network $\gG$ gives $O(|\gE| \cdot F^2 + |\gE| \cdot D)$ assuming GCN + Inner Product (b) optimizing over diffusion episodes is $O(|\sT| \cdot D \cdot N)$ where $F$ is the maximum layer dimension in $f_{\textsc{enc}}$. 
The overall complexity per iteration is $O(|\gE| \cdot F^2 + |\gE| \cdot D + |\sT| \cdot D \cdot N)$.

\section{Experiments}
\newcolumntype{K}[1]{>{\centering\arraybackslash}p{#1}}
\newcolumntype{Y}{>{\raggedleft\arraybackslash}X}

In this section, we present our experimental results on multiple datasets from real-world social networks and public Stack-Exchanges\footnote{\url{https://archive.org/details/stackexchange}}. We examine two popular social networks Digg and Weibo. %
\begin{itemize}[leftmargin=*]
\item \textbf{Digg}~\cite{digg}: A social platform where users vote on news stories.
The sequence of votes on each story constitutes a diffusion cascade, while the social network comprises friendship links among voters.
We retain only users who have voted on at least 40 stories.

\item \textbf{Weibo}~\cite{locality}: A Chinese micro-blogging platform,
where the social network consists of follower links, and cascades reflect re-tweeting behavior. We choose the 5000 most popular users. %

\end{itemize}
\textbf{Stack-Exchanges:} Community Q\&A websites where users post questions and answers on a wide range of topics.
The inter-user knowledge-exchanges on various interaction channels (\textit{e.g.}, question, answer, comment, upvote, etc.), constitute the social network.
Cascades correspond to chronologically ordered series of posts associated with the same tag, \textit{e.g.}, ``google-pixel-2" on Android.
We choose three Stack-Exchanges, Android, Christianity and Travel, spanning diverse themes.
Dataset statistics are provided in Table~\ref{tab:dataset_stats}.

\subsection{Baselines}
We compare~\name~against state-of-the-art representation learning methods for diffusion prediction since they have been shown to significantly outperform classical models (\textit{e.g.}, IC and LT)~\cite{topolstm, inf2vec}.

\begin{itemize}[leftmargin=*]
    \item \textbf{CDK}~\cite{cdk}: an embedding method that models information spread as a heat diffusion process in the representation space of users.
    \item \textbf{Emb-IC}~\cite{wsdm16}: an embedded cascade model that generalizes IC to learn user representations from partial orders of user activations.
    \item \textbf{Inf2vec}~\cite{inf2vec}: an influence embedding method that combines local propagation structure and user co-occurrence in cascades. 
    \item \textbf{DeepDiffuse}~\cite{deepdiffuse}: an attention-based RNN that operates on just the sequence of previously influenced users, to predict diffusion.
    \item \textbf{CYAN-RNN}~\cite{cyanrnn}: a sequence-based RNN that uses an attention mechanism to capture cross-dependence among seed users.
    \item \textbf{SNIDSA}~\cite{cikm18_attention}: an RNN-based model to compute structure attention over the local propagation structure of a cascade.    
    \item \textbf{Topo-LSTM}~\cite{topolstm}: a recurrent model that exploits the local propagation structure of a cascade through a dynamic DAG-LSTM.
\end{itemize}

\begin{table}[t]
\centering
\small
\begin{tabular}{@{}p{0.24\linewidth}K{0.08\linewidth}K{0.08\linewidth}K{0.11\linewidth}K{0.14\linewidth}K{0.13\linewidth}@{}}
\toprule
\multirow{2}{*} & \multicolumn{2}{c}{\small \textbf{Social Networks}}  & \multicolumn{3}{c}{\small \textbf{Stack-Exchange Networks}}\\
\cmidrule(lr){2-3} \cmidrule(lr){4-6}
\textbf{Dataset} & \small \textbf{Digg}  & \small \textbf{Weibo} & \small \textbf{Android} & \small\textbf{Christianity}  & \small\textbf{Travel}\\
\midrule
\textbf{\small \# Users} & 8,602 & 5,000 & 9,958 & 2,897 & 8,726\\
\textbf{\small \# Links} & 173,489 & 123,691 & 48,573 &  35,624 &  76,555\\
\textbf{\small \# Cascades} & 968 & 23,475 & 679 &  589 & 711\\
\textbf{\small Avg. cascade len} & 100.0 & 23.6 & 33.3 &  22.9 & 26.8\\
\bottomrule
\end{tabular}
\caption{Statistics of datasets used in our experiments}
\label{tab:dataset_stats}
\vspace{-20pt}
\end{table}

\newcommand*{\factor}{0.039}
\begin{table*}[t]
\centering
\small
\begin{tabular}{@{}p{0.11\linewidth}K{\factor\linewidth}K{\factor\linewidth}K{\factor\linewidth}K{\factor\linewidth}K{\factor\linewidth}K{\factor\linewidth}K{\factor\linewidth}K{\factor\linewidth}K{\factor\linewidth}K{\factor\linewidth}K{\factor\linewidth}K{\factor\linewidth}K{\factor\linewidth}K{\factor\linewidth}K{\factor\linewidth}@{}} \\
\toprule
\multirow{1}{*}{\textbf{Method}} & \multicolumn{3}{c}{\textbf{Digg}}  &   \multicolumn{3}{c}{\textbf{Weibo} } & \multicolumn{3}{c}{\textbf{Android}}  & \multicolumn{3}{c}{\textbf{Christianity}} & \multicolumn{3}{c}{\textbf{Travel}} \\
\cmidrule(lr){2-4} \cmidrule(lr){5-7} \cmidrule(lr){8-10} \cmidrule(lr){11-13} \cmidrule(lr){14-16}
\multirow{1}{*}{\textbf{MAP}} & \textbf{@10} & \textbf{@50} & \textbf{@100} & \textbf{@10} & \textbf{@50} & \textbf{@100} & \textbf{@10} & \textbf{@50} & \textbf{@100}& \textbf{@10} & \textbf{@50} & \textbf{@100}& \textbf{@10} & \textbf{@50} & \textbf{@100}\\
\midrule
\textbf{CDK} &  0.0437 & 0.0222 & 0.0228 & 0.0130 & 0.0106 & 0.0123 & 0.0319 & 0.0121 & 0.0125 & 0.0876 & 0.0531 & 0.0578 & 0.0650 & 0.0333 & 0.0341 \\
\textbf{Emb-IC} & 0.0862 & 0.0431 & 0.0431 & 0.0140 & 0.0116 & 0.0131 & 0.0505 & 0.0248 & 0.0267 & 0.1340 & 0.0905 & 0.0962 & 0.0924 & 0.0584 & 0.0609 \\
\textbf{Inf2vec} & 0.1189 & 0.0554 & 0.0546 & 0.0156 & 0.0103 & 0.0121 & 0.0412 & 0.0141 & 0.0150 & 0.1824 & 0.0790 & 0.0852 & 0.1245 & 0.0495 & 0.0529\\
\textbf{DeepDiffuse} & 0.0919 & 0.0460 & 0.0471 & 0.0291 & 0.0186 & 0.0213 & 0.0437 & 0.0228 & 0.0250 & 0.1632 & 0.0828 & 0.0831 & 0.1220 & 0.0675 & 0.0693  \\
\textbf{CYAN-RNN} & 0.1188 & 0.0479 & 0.0427 & 0.0296 & 0.0207 & 0.0234 & 0.0520 & 0.0276 & 0.0296 & 0.1971 & 0.1229 & 0.1304 & 0.1551 & 0.0791 & 0.0799\\
\textbf{SNIDSA} & 0.0941 & 0.0363 & 0.0348 & 0.0224 & 0.0146 & 0.0169 & 0.0397 & 0.0207 & 0.0222 & 0.1233 & 0.0699 & 0.0781 & 0.0857 & 0.0562 & 0.0585   \\
\textbf{Topo-LSTM} & 0.1193 & 0.0577 & 0.0587 & 0.0325 & 0.0226 & 0.0247 & 0.0595 & 0.0283 & 0.0289 & 0.1811 & 0.0989 & 0.0991 & 0.1393 & 0.0773 & 0.0783 \\
\midrule
\textbf{Inf-VAE+MLP} & 0.1587 & 0.0774 & 0.0719 & 0.0322 & 0.0211 & 0.0234 & 0.0584 & 0.0272 & 0.0285 & 0.2549 & 0.1355 & 0.1402 & 0.1865 & 0.0897 & \textbf{0.0913} \\
\textbf{Inf-VAE+GCN} & \textbf{0.1642} & \textbf{0.0779} & \textbf{0.0724} & \textbf{0.0373} & \textbf{0.0230} & \textbf{0.0257} & \textbf{0.0601} & \textbf{0.0290} & \textbf{0.0304} & \textbf{0.2594} & \textbf{0.1413} & \textbf{0.1461} & \textbf{0.1924} & \textbf{0.0906} & 0.0910 \\
\bottomrule
\end{tabular}
\caption{Experimental results for diffusion prediction on 5 datasets ($MAP@K$ scores for $K = 10, 50$ and $100$), the \textit{seed set percentage} varies in the range to 10 to 50\% users in each test cascade. 22\% relative gains in MAP@10 (on average) over the best baseline.}
\label{tab:map_results}
\vspace{-18pt}
\end{table*}

\subsection{Experimental Setup}
We denote our two model variants with GCN and MLP  architectures, by ~\name+\textsc{GCN} and ~\name+\textsc{MLP} respectively.
We randomly sample 70\% of the cascades for training, 10\% for validation and remaining 20\% for testing.
We consider the task of predicting the set of all influenced users as a retrieval problem~\cite{topolstm,wsdm16,cyanrnn,inf2vec}.
The fraction of users sampled from each test cascade to serve as the seed set is defined as \textit{\textbf{seed set percentage}}, which is varied from 10\% to 50\% to create a large evaluation test-bed spanning diverse cascade lengths.
The likelihood of influencing an inactive user determines its rank (Eqn.~\ref{eqn:user_prediction}).
We use MAP$@K$ (Mean Average Precision) and Recall$@K$ as evaluation metrics. Note that MAP$@K$ considers both the \textit{existence} and \textit{position} of ground-truth target users in the rank list, while Recall$@K$ only reports occurrence within top-$K$ ranks.

Hyper-parameters are tuned by evaluating MAP@10 on the validation set.
Since Emb-IC generalizes IC, we use 1000 Monte Carlo simulations to estimate influence probabilities.
Since the recurrent neural models (\textit{e.g.,} Topo-LSTM) are trained for next user prediction, 
we use the ranking induced by user activation probabilities for diffusion prediction, which we found to significantly outperform a similar simulation approach.
For Inf2vec, we examine several seed influence aggregation functions (Ave, Sum, Max, and Latest) to report the best results.
Our reported results are averaged over 10 independent runs with different random weight initializations. Our implementation of~\name~is publicly available\footnote{\url{https://github.com/aravindsankar28/Inf-VAE}}.

\subsection{Experimental Results}
We note the following key observations from our experimental results comparing~\name~against competing baselines (Table~\ref{tab:map_results}).

Methods that do not explicitly model sequential activation orders (\textit{e.g.}, CDK and Emb-IC), perform markedly worse than their counterparts.
Modeling local projected cascade structures with neural recurrent models results in improvements (\textit{e.g.}, Topo-LSTM and others).
Jointly modeling social homophily derived from global network structure and temporal influence by our model 
~\name~yields significant relative gains of 22\% ($MAP@10$) on average across all datasets.
~\name+GCN consistently beats the MLP variant, validating the power of graph convolutional networks in effectively propagating higher-order local neighborhood features.

\begin{figure*}[t]
    \vspace{-6pt}
    \centering
    \includegraphics[width=\linewidth]{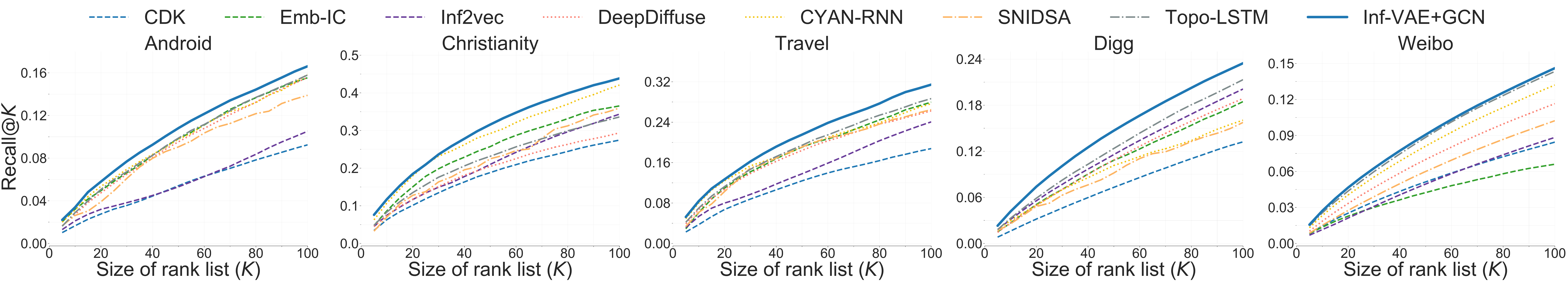}
    \vspace{-22pt}
    \caption{Experimental results for diffusion prediction on 5 datasets, Recall$@K$ scores on varying size of the rank list $K$}
    \label{fig:recall}
    \vspace{-10pt}
\end{figure*}

Figure~\ref{fig:recall} depicts the variation in recall with size of rank list $K$.
As expected, recall increases with $K$, however, the relative differences across methods is much smaller.
~\name~consistently outperforms baselines across a wide range of $K$ values.
For instance, 
the Christianity dataset has seed sets with 2-10 users, and corresponding target sets with 10-15 users out of a possible 3000.
Here, a recall$@100$ of 0.45 for~\name~is quite impressive, especially considering the absence of explicit re-share links and the noise associated with real-world diffusion processes.
We restrict our remaining analyses to~\name+GCN~since it consistently beats the MLP variant.

\begin{figure}[t]
    \centering
    \vspace{-6pt}
    \includegraphics[trim={0 0 0 0.8cm},clip,width=\linewidth]{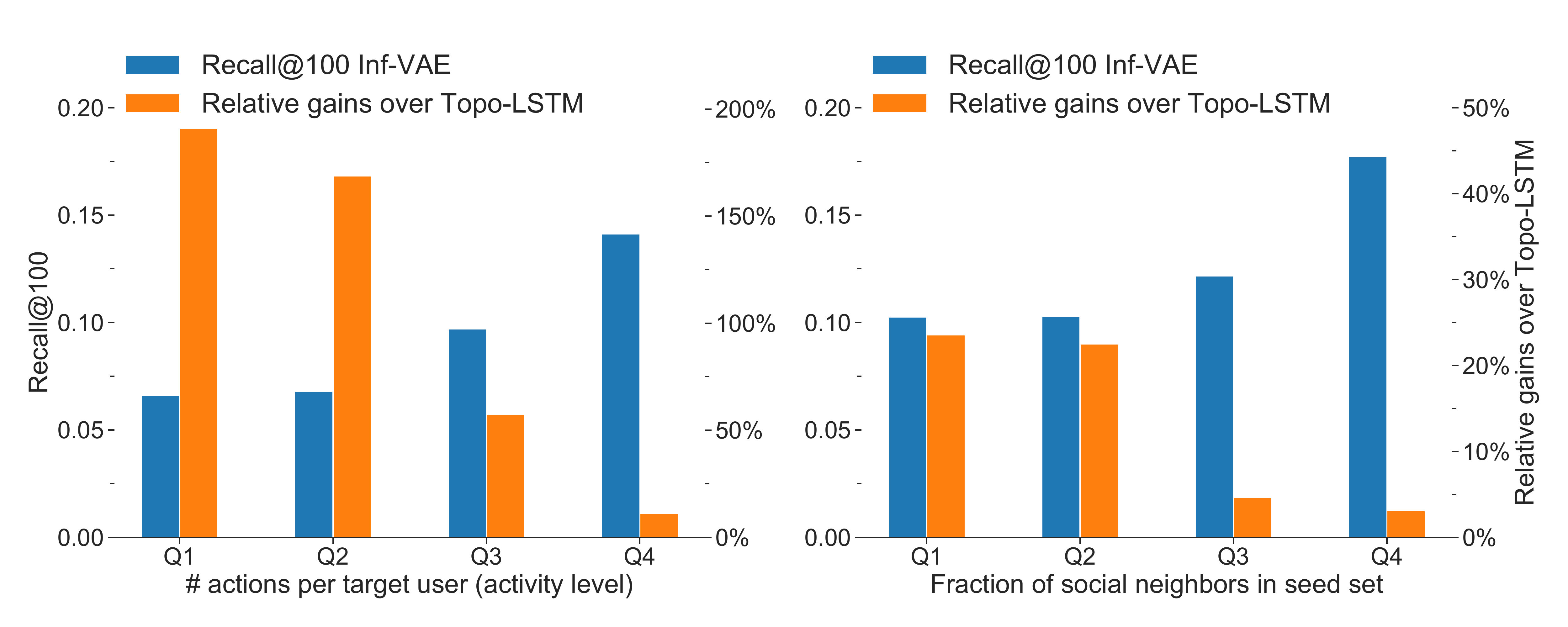}
    \vspace{-22pt}
    \caption{Performance across user quartiles on \textit{diffusion activity level}, and \textit{seed neighbor fraction} (Q1: lowest, Q4: highest).~\name~has higher gains for users with sparse activities and lacking direct neighbors in seed sets (quartiles Q1-Q3).}
    \vspace{-16pt}
    \label{fig:social_activity}
\end{figure}

\subsection{Impact of Social and Behavior Sparsity}
In this section, we analyze the benefits of explicitly modeling social homophily through VAEs, in comparison to the best baseline (Topo-LSTM) that only considers local propagation structures.
\begin{itemize}[leftmargin=*]
\item \textbf{{Users with sparse diffusion activities.}}
We divide users into quartiles by their \textit{activity levels}, which is the number of participating cascades per user. 
We evaluate \textit{target recall}$@100$ for each user $u$, defined as the fraction of times $u$ was predicted correctly within top-$100$ ranks.
In Figure \ref{fig:social_activity}(a), we depict 
both recall scores and relative gains (over Topo-LSTM) across activity quartiles.

While target recall increases with activity levels,~\name~significantly improves performance for inactive users (quartiles Q1-Q3).
Thus, modeling social homophily through VAEs contributes to massive gains for users with \textit{sparse diffusion activities}.
Interestingly, Topo-LSTM performs comparably on the most active users (quartile Q4), which indicates the potential of purely local sequential modeling techniques for highly active users.

\item  \textbf{Users that lack direct social neighbors in seed sets.}
We separate users into quartiles by \textit{seed neighbor fraction}, which is computed as the fraction of seed users that are direct social neighbors, averaged over the training examples. We similarly report target recall$@K$ and relative gains across quartiles (Figure~\ref{fig:social_activity}(b)).

As expected, performance increases with seed neighbor fraction. Note higher relative gains over Topo-LSTM for users that lack direct neighbors in the seed set (quartiles Q1-Q3). 
This demonstrates the ability of~\name~to implicitly regularize seed user representations based on higher-order social neighborhoods captured by GCN-based autoencoders.
Again, we find that local sequential models suffice for users with 
large seed neighbor fractions, as evidenced by the results of Topo-LSTM in quartile Q4.

\end{itemize}

\renewcommand*{\factor}{0.071}
\begin{table}[H]
\vspace{-14pt}
\centering
\small
\begin{tabular}{@{}p{0.31\linewidth}K{\factor\linewidth}K{\factor\linewidth}K{\factor\linewidth}K{\factor\linewidth}K{\factor\linewidth}K{\factor\linewidth}K{\factor\linewidth}K{\factor\linewidth}@{}} \\
\toprule
{\textbf{Metric}} &  \multicolumn{3}{c}{\textbf{Weibo} } & \multicolumn{3}{c}{\textbf{Android}} \\
\cmidrule(lr){2-4} \cmidrule(lr){5-7}
MAP & @10 & @50 & @100 & @10 & @50 & @100 \\
\midrule
(0) Default  & \textbf{0.0373} & \textbf{0.0230} & \textbf{0.0257} & \textbf{0.0601} & \textbf{0.0290} & \textbf{0.0304}\\
(1)$\mV_S = \mV_R \not \independent \mZ$ & 0.0353 & 0.0220 & 0.0248 & 0.0558 & 0.0275 & 0.0287\\
(2)$\mV_S \independent \mZ$ & 0.0351 & 0.0213 & 0.0240 & 0.0595 & 0.0285 & 0.0301\\
(3)$\mV_R \independent \mZ$ & 0.0326 & 0.0217 & 0.0241 & 0.0567 & 0.0276 & 0.0291\\
(4)$\mV_S \independent \mZ, \mV_R \independent \mZ$ & 0.0313 & 0.0205 & 0.0235 & 0.0542 & 0.0274 & 0.0289\\
\midrule
(5) Remove Coattention & 0.0307 & 0.0207 & 0.0233 & 0.0553 & 0.0270 & 0.0284 \\ 
(6) Separate Attentions & 0.0293 & 0.0217 & 0.0192 & 0.0570 & 0.0277 & 0.0291 \\
\midrule
(7) Static-Pretrain & 0.0342 & 0.0203 & 0.0226 & 0.0606 & 0.0281 & 0.0292 \\
\bottomrule
\end{tabular}
\caption{Ablation study on architecture design ($MAP@K$ scores for $K = 10, 50, 100$), $\independent$ denotes variable independence}
\vspace{-20pt}
\label{tab:ablation_results}
\end{table}

\subsection{Ablation Study and Sensitivity Analysis}
In this section, we first present an \textit{ablation study} followed by a sensitivity analysis on \textit{seed set percentage} and \textit{hyper-parameters}.
\vspace{-2pt}

\subsubsection{\textbf{Ablation Study}}
We analyze model design choices including homophily via VAEs and co-attention, in Android and Weibo.

\vspace{2pt}

\noindent \textbf{Social Homophily:} We examine ways to condition $\mV_S$, $\mV_R$ on $\mZ$:
\vspace{-2pt}
\begin{enumerate}[leftmargin=*]
    \item $\mV_S$ and $\mV_R$ are identical and are conditioned on $\mZ$ through hyper-parameter $\lambda_s (=\lambda_r)$, \textit{i.e.}, $\mV_S = \mV_R \not \independent \mZ$ (note that this is different from setting $\lambda_s = \lambda_r$ without enforcing $\mV_S=\mV_R$).
    \item $\mV_S$ is a free variable conditionally independent of $\mZ$, \textit{i.e.}, $\mV_S \independent \mZ$, which is equivalent to setting $\lambda_s=0$.
    \item $\mV_R$ is a free variable, \textit{i.e.}, $\mV_R \independent \mZ$, which is the inverse of (3).
    \item $\mV_S$ and $\mV_R$ are both free variables conditionally independent of $\mZ$ ($\lambda_s= \lambda_r =0$), \textit{i.e.}, $\mV_S \independent \mZ, \mV_R \independent \mZ$.
\end{enumerate}
\vspace{-2pt}

\noindent Independent conditioning of $\mV_S$ and $\mV_R$ on $\mZ$ (default) achieves best results. %
Enforcing $\mV_S=\mV_R$ (row 1) is clearly inferior,
which validates the choice of differentiating user roles.
Notably, allowing $\mV_S$ to be a free variable 
results in minor performance degradation (row 2), while the drop is significant when $\mV_R$ is independent of $\mZ$ (row 3).
As expected, setting both $\mV_S$ and $\mV_R$ as free variables (row 4),
performs the worst due to lack of social homophily signals.

\vspace{2pt}
\noindent \textbf{Co-attention:}
We conduct two ablation studies defined by:
\vspace{-3pt}
\begin{enumerate}[leftmargin=*]
\setcounter{enumi}{4}
    \item Replace co-attention with meanpool over concatenated sender and temporal influence vectors, followed by a dense layer.
    \item Replace co-attention with two separate attentions on the sender and temporal influence sequences, followed by concatenation.
\end{enumerate}
\vspace{-3pt}
Learning co-attentional weights (default) consistently outperforms mean pooling (5), illustrating the benefits of assigning variable contributions to seed users. 
Using separate attentions (6) significantly deteriorates results, which indicates 
the existence of complex non-linear correlations between the social and temporal latent factors.

\vspace{2pt}
\noindent \textbf{Joint Training:} In (2), we replace joint block-coordinate optimization (Alg~\ref{alg:opt}) with a single step over cascades with pre-trained user embeddings (line 2), \textit{i.e.}, $\mZ$ is not updated based on cascades.

\noindent 
Joint training is beneficial when social interactions are noisy (\textit{e.g.,} Weibo) in comparison to focused stack-exchanges such as Android.

\begin{figure}[t]
    \centering
    \includegraphics[width=\linewidth]{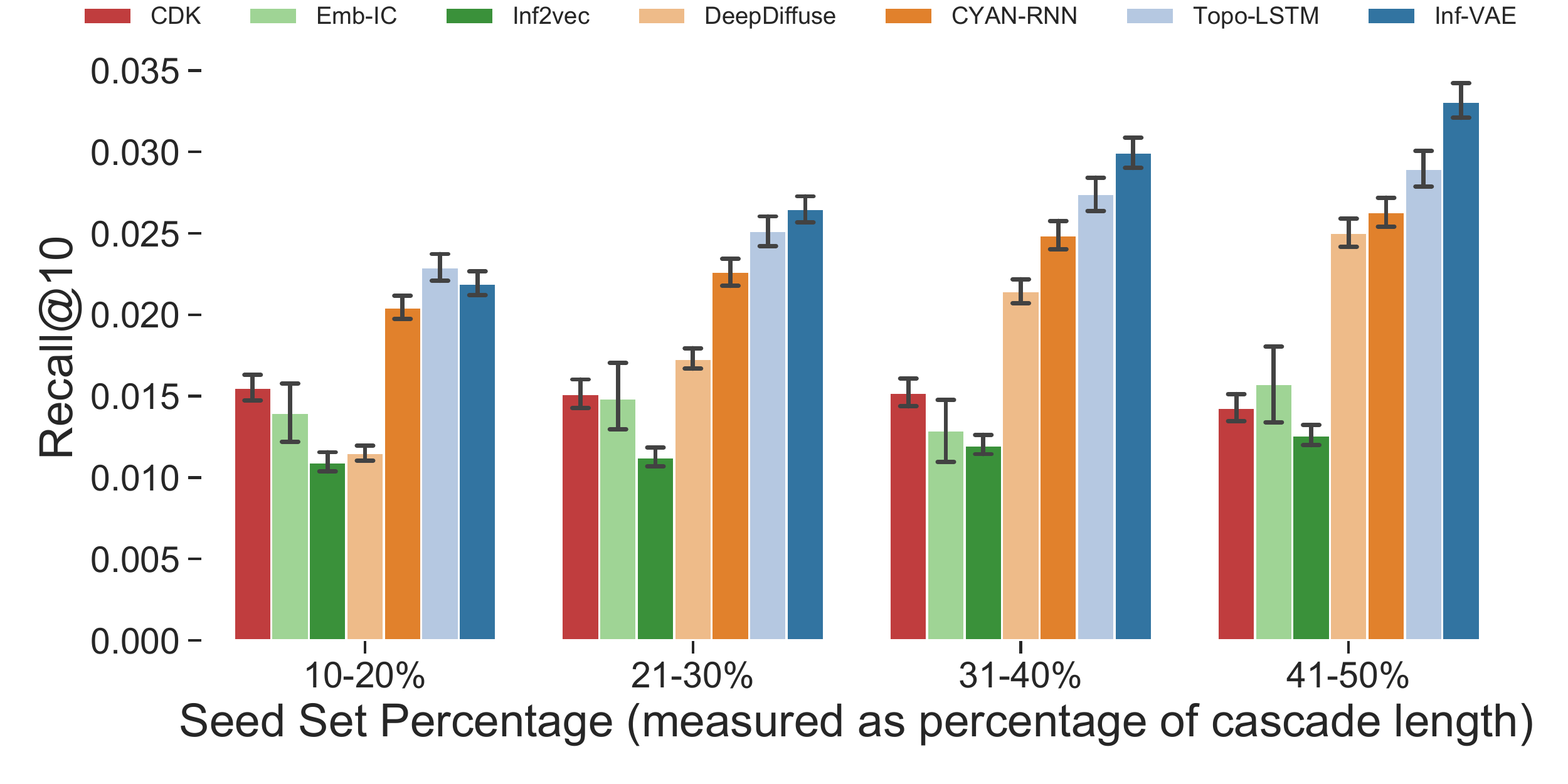}
    \vspace{-25pt}
    \caption{Impact of seed set percentage in Weibo.~\name~achieves higher gains 
    for larger seed set fractions.}
    \label{fig:seed_analysis}
    \vspace{-13pt}
\end{figure}

\subsubsection{\textbf{Impact of Seed Set Percentage.}}
We divide the test set into quartiles based on \textit{seed set percentage}, and report performance per quartile.
Since we require a sizable number of test examples per quartile to obtain unbiased estimates, we use the Weibo dataset. %

Figure~\ref{fig:seed_analysis} depicts Recall$@10$ scores in different ranges. %
First, recall scores increase with seed set percentage since larger seed sets enable better model predictions; and target set size reduces with increase in seed set percentage.
Second, relative gains of~\name~over baselines increase with seed set percentage. %
This highlights the capability of co-attention in focusing on relevant users based on both social homophily and temporal influence factors.

\subsubsection{\textbf{Impact of $\lambda_s$ and $\lambda_r$}}
\begin{figure}[t]
\vspace{-5pt}
    \centering
    \includegraphics[width=\linewidth]{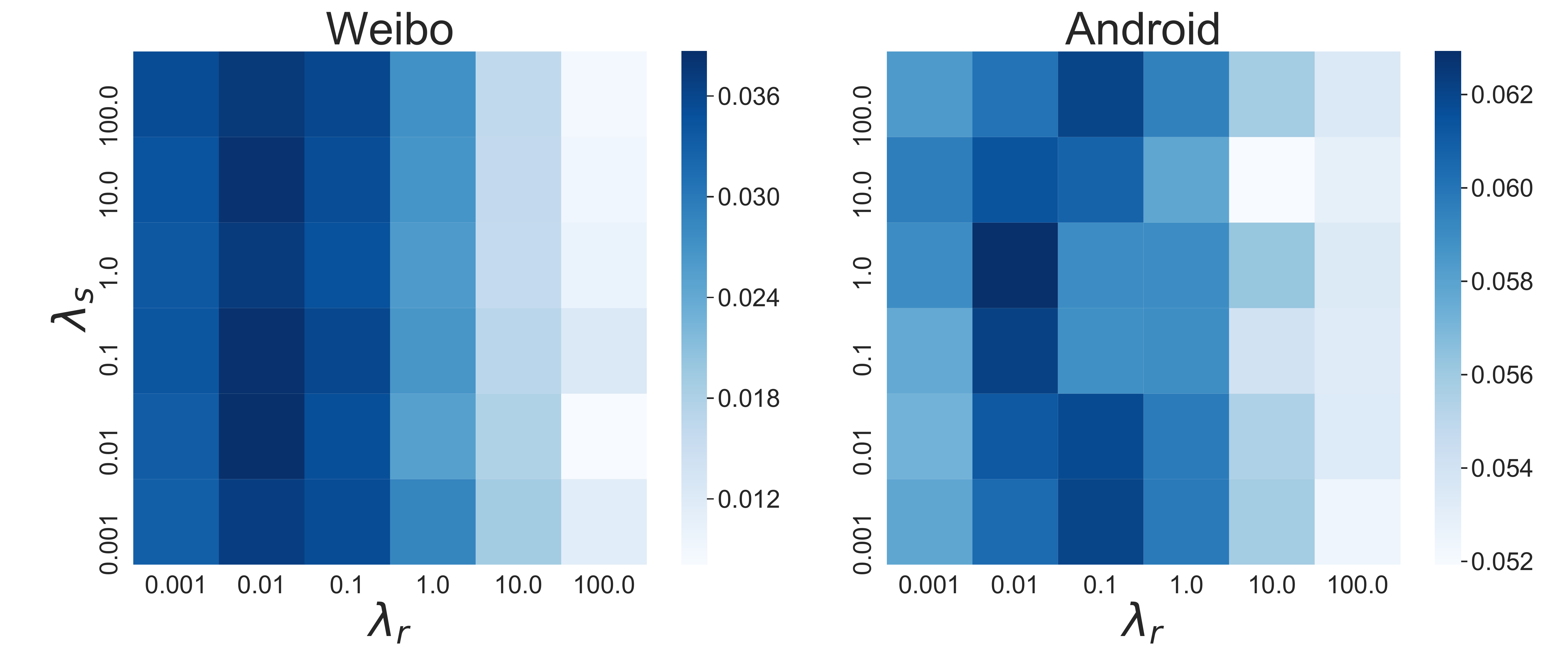}
     \vspace{-18pt}
    \caption{$MAP@10$ on varying $\lambda_s$, $\lambda_r$ over Android and Weibo. Performance is more sensitive to variations in $\lambda_r$ than $\lambda_s$.}
    \vspace{-12pt}
    \label{fig:lambda_heatmap}
\end{figure}

Hyper-parameters $\lambda_s$ and $\lambda_r$ control the degree of dependence of the sender and receiver variables $\mV_S, \mV_R$ on the social variables $\mZ$.
Figure~\ref{fig:lambda_heatmap} depicts performance ($MAP@10$) on Android and Weibo datasets.
The performance is sensitive to variations in $\lambda_r$ with best values around 0.01 and 0.1,
while $\lambda_s$ results in minimal variations.
Furthermore, the best values of $\lambda_s, \lambda_r$ are stable in a broad range of values that transfer across datasets, indicating that~\name~requires minimal tuning in practice.
Since $\lambda_p$ has minimal performance impact, we exclude it from our analysis.

\begin{figure}[t]
    \vspace{-3pt}
    \centering
        \subfigure[b][Runtime on Weibo and Android
        ]{
        \centering
        \includegraphics[width=0.45\linewidth]{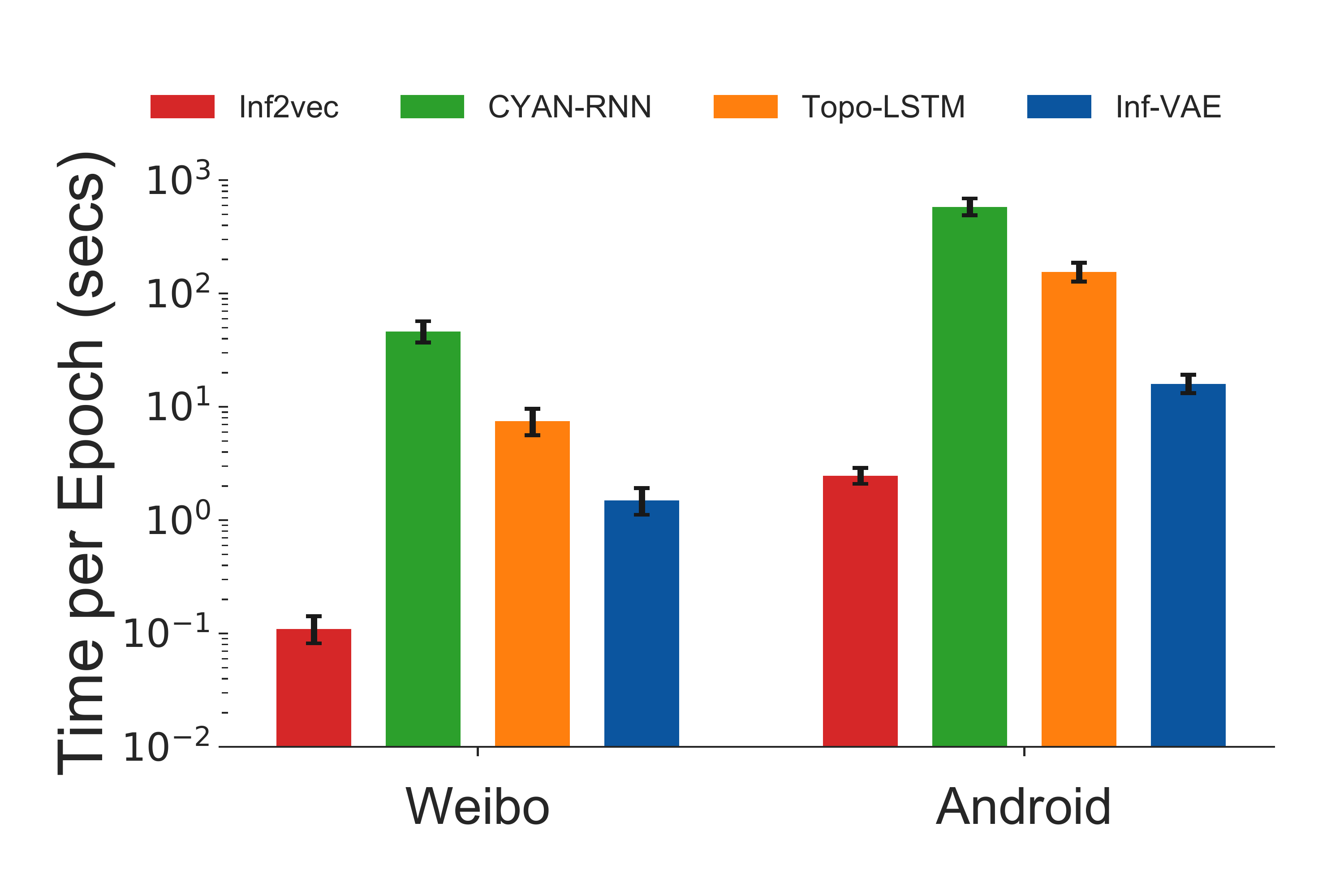}
        \label{fig:runtime}
        }
        \hspace{2pt}
        \subfigure[b][Scalability on synthetic dataset]{
        \centering
        \includegraphics[width=0.45\linewidth]{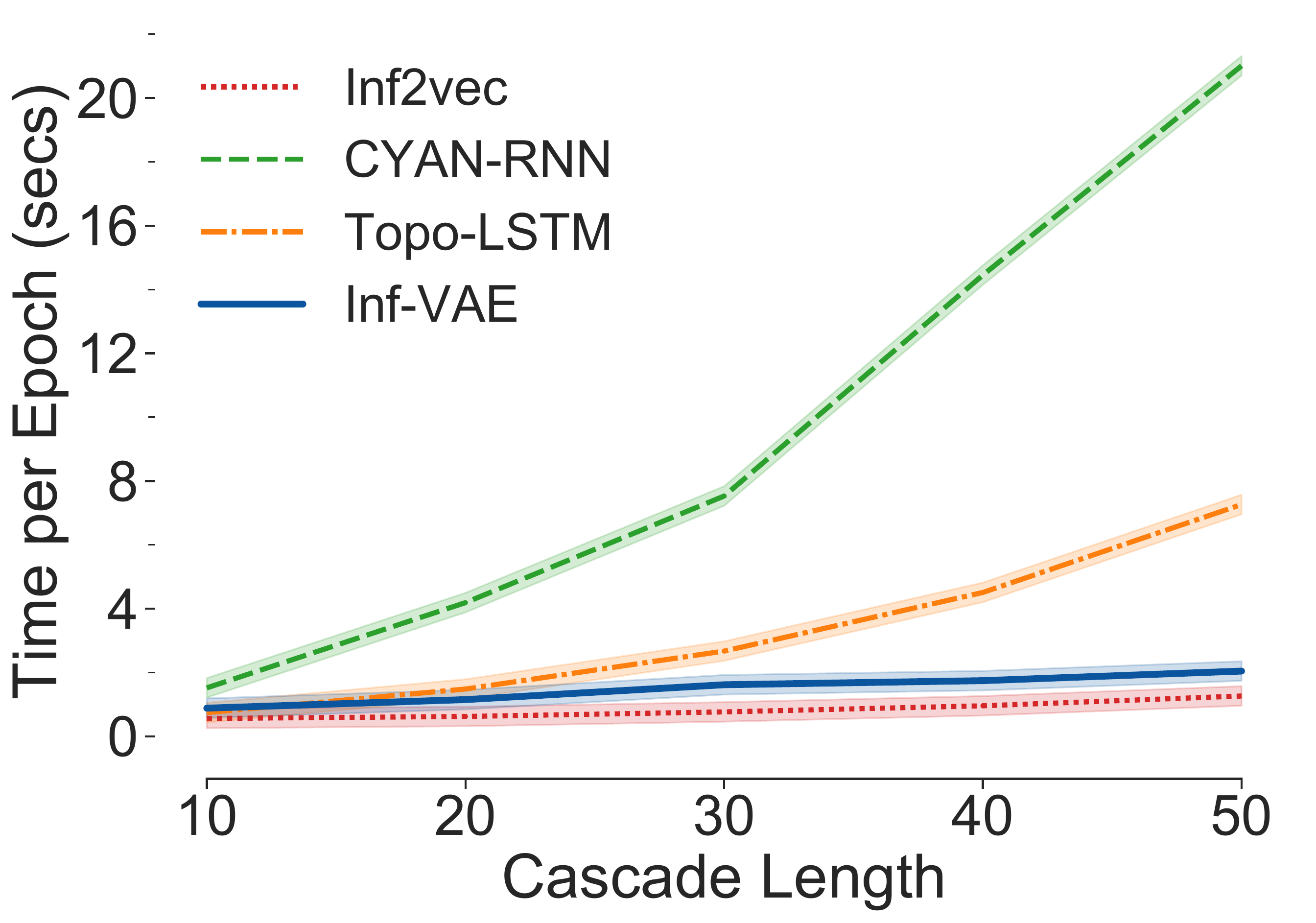}
        \label{fig:scalability} }        
    \vspace{-10pt}
    \caption{Running time and scalability comparison of~\name~with several baselines.~\name~is faster than recurrent models (Topo-LSTM, CYANRNN) by an order of magnitude.}
    \vspace{-11pt}
\end{figure}
\subsubsection{\textbf{Runtime Analysis}.} 
In our experiments, all methods converge within 50 epochs with similar convergence rates.
For the sake of brevity, we only compare runtime per epoch, which includes one step over the social network and cascades for~\name.

From figure~\ref{fig:runtime}, Inf2vec is the fastest while~\name~ comes second.
Thus,~\name~achieves a good trade-off between 
expensive recurrent models (\textit{e.g.}, Topo-LSTM) and simpler embedding methods (\textit{e.g.}, Inf2vec), with consistently superior results.
\subsubsection{\textbf{Scalability Analysis}.} We analyze scalability on cascade sequences of varying lengths.
Since real-world datasets possess heavily biased length distributions, we instead 
synthetically generate a 
Barabasi-Albert~\cite{barabasi} network of 2000 users and simulate diffusion cascades using an IC model.
We compare training times per epoch for each cascade length ($l$) in the range of 10 to 50.

Figure~\ref{fig:scalability} depicts
linear scaling for~\name~and Inf2vec \textit{wrt} cascade length.
Recurrent methods scale poorly due to the  sequential nature of back-propagation through time (BPTT), resulting in prohibitive costs for long cascade sequences.
On the other hand,~\name~avoids BPTT through efficient parallelizable co-attentions.

\section{Conclusion}
In this paper, we present a novel variational autoencoder framework (\name) to jointly embed homophily and influence in diffusion prediction.
Given a sequence of seed user activations,~\name~employs an expressive co-attentive fusion mechanism to jointly attend over their social and temporal variables, capturing complex correlations.
Our experimental results on two social networks and three stack-exchanges indicate significant gains over state-of-the-art methods.

In future, ~\name~can be extended to include multi-faceted user attributes owing to the generalizable nature of our VAE framework. While the current implementation employs GCN networks, we foresee direct extensions with neighborhood sampling~\cite{graphsage} to enable scalability to social networks with millions of users. 
We also plan to explore neural point processes to predict user activation times. 
Finally, similar frameworks may be examined for joint temporal co-evolution of social network and diffusion cascades.

\section{Acknowledgements}
Research was sponsored in part by U.S. Army Research Lab. under Cooperative Agreement No. W911NF-09-2-0053 (NSCTA), DARPA under Agreements No. W911NF-17-C-0099 and FA8750-19-2-1004, National Science Foundation IIS 16-18481, IIS 17-04532, and IIS-17-41317, and DTRA HDTRA11810026.

\bibliographystyle{ACM-Reference-Format}
\bibliography{diffusion}

\end{document}


\title[Integrating Social Homophily and Temporal Influence in Diffusion Prediction]{A Deep Generative Approach to Integrate Social Homophily and Temporal Influence in Diffusion Prediction \\ Supplementary Material\vspace{-5pt}}

\author{Ben Trovato}
\authornote{Dr.~Trovato insisted his name be first.}
\orcid{1234-5678-9012}
\affiliation{%
  \institution{Institute for Clarity in Documentation}
  \streetaddress{P.O. Box 1212}
  \city{Dublin}
  \state{Ohio}
  \postcode{43017-6221}
}
\email{trovato@corporation.com}

\renewcommand{\shortauthors}{B. Trovato et al.}
\renewcommand{\algorithmicrequire}{\textbf{Input:}}
\renewcommand{\algorithmicensure}{\textbf{Output:}}
\renewcommand{\citep}{\cite}

\makeatletter
\newcommand{\algmargin}{\the\ALG@thistlm}
\makeatother

\maketitle

\input{./texfiles/appendix.tex}
\bibliographystyleNew{ACM-Reference-Format}
\bibliographyNew{diffusion}